# Transient Behavior near Liquid-Gas Interface at Supercritical Pressure

Jordi Poblador-Ibanez[1,*], William A. Sirignano[2]

*University of California, Irvine, CA 92697-3975, United States*

**Abstract**

Numerical heat and mass transfer analysis of a configuration where a cool liquid hydrocarbon is suddenly introduced to a hotter gas at supercritical pressure shows that a well-defined phase equilibrium can be established before substantial growth of typical hydrodynamic instabilities. The equilibrium values at the interface quickly reach near-steady values. Sufficiently thick diffusion layers form quickly around the liquid-gas interface (e.g., 3-10 $\mu$m for the liquid phase and 10-30 $\mu$m for the gas phase in 10-100 $\mu$s), where density variations become increasingly important with pressure as mixing of species is enhanced. While the hydrocarbon vaporizes and the gas condenses for all analyzed pressures, the net mass flux across the interface reverses as pressure is increased, showing that a clear vaporization-driven problem at low pressures may present condensation at higher pressures. This is achieved while heat still conducts from gas to liquid. Analysis of fundamental thermodynamic laws on a fixed-mass element containing the diffusion layers proves the thermodynamic viability of the obtained results.

## 1. Introduction

Liquid fuel injectors are present in many combustion applications. These injectors are designed to optimize the liquid breakup process or atomization, where droplets form and then vaporize and mix with the surrounding gas (i.e., oxidizer), allowing the combustion chemical reaction to occur. Therefore, understanding how this liquid disruption process develops is crucial to obtain a good performance of combustion chambers and engineering applications (e.g., diesel engines, rocket engines).

These combustion chambers often operate at very high pressures, since combustion efficiency is improved and a higher specific energy conversion can be obtained. These pressures can be above the critical pressure of the injected fuel. When this supercritical situation occurs, the thermodynamics and fluid dynamics of the injection phenomena are drastically modified. Since gas readily dissolves in the liquid at high pressures, a new solution is formed typically with a much higher critical pressure than the original liquid. At subcritical pressures, a clear distinction between the liquid and the gas phase exists and the liquid breakup is mainly driven by capillary forces and hydrodynamic instabilities, depending on the fluid properties and configuration of the problem. However, at high-pressure regimes close or above the critical pressure of the new liquid solution, diffusion can become one of the main drivers of the mixing process and a well-defined liquid-gas interface may not be easily identified [1].

---
[*]Corresponding author
  *Email address:* poblador@uci.edu (Jordi Poblador-Ibanez)
  [1]Graduate Student Researcher, Department of Mechanical and Aerospace Engineering.
  [2]Professor, Department of Mechanical and Aerospace Engineering.

In fact, liquid injected into an environment exceeding its critical properties will undergo a transition to a supercritical thermodynamic state. In this process, the liquid goes through a near-critical state where liquid-like densities and gas-like diffusivities will be present. Consequently, the classical liquid-gas interface seems to disappear even before supercritical conditions are achieved. Furthermore, as the critical point is reached, the liquid-gas density ratio reduces and surface tension approaches zero, increasing the aerodynamic effects on the liquid breakup [2, 3]. This situation has often been described with the wrong assumption that the liquid fuel simply transitions to a supercritical gas-like state in a single-phase manner. Experimental results from Mayer et al. [4] show the behavior, but no emphasis is made on whether the observed results correspond to an actual transition of the injected liquid to a supercritical gas-like state or if rapid disruption of the liquid stream due to enhanced aerodynamic effects cause the formation of small droplets, which in turn could not be easily seen in experimental setups.

Recent works from Jarrahbashi et al. [5, 6], and Zandian et al. [7, 8] present results for the liquid disruption process of incompressible round jets and planar sheets, respectively, when injected into a gas. Simulations with varying Reynolds number and Weber number show that high-speed flows not only can induce a fast breakup process generating small droplets, but also similar densities for the liquid and gas phases (i.e., high-density or high-pressure environments) and reduced surface tension can cause a similar behavior. In this scenario, the formation of certain structures is slowed down in favor of another formation process for droplets and an enhanced radial development of the two-phase mixture. These works do not include any treatment of the energy equation or species mixing, but show results likely to be seen in supercritical pressure injection. These results suggest that when the gas-like appearance of the jet is observed, it may still be in a liquid state, instead of a supercritical state of the fuel. Therefore, it becomes necessary to determine the characteristic times of each process (i.e., analyze whether the transition to a supercritical state occurs faster than the liquid disruption cascade or not).

On the other hand, several chemical species are present in real combustion applications, thus modifying completely the fluid behavior. Under supercritical pressure conditions, not only the mixing of vaporized fuel and surrounding gas must be accounted for, but also the gas dissolving into the liquid phase due to phase-equilibrium conditions. This diffusion of species generates a liquid solution and a gas mixture with varying composition around the liquid-gas interface which modifies the fluid properties. That is, mixture critical properties will differ from pure species critical properties. Precisely, critical pressure of the liquid solution will increase close to the interface, even above the chamber pressure in some configurations. In this situation, the transition of the injected liquid to a supercritical state is delayed and a distinguishable two-phase behavior can be maintained. Some experimental results also provided in [4] show that, for configurations including different species, the liquid-gas interface can be restored depending on the local mixture composition, temperature and pressure. Moreover, numerical results provided by Jordà-Juanós and Sirignano [9] show that phase-equilibrium laws together with a real-gas equation of state are able to reproduce the coexistence of two phases at supercritical pressures.

Thus, the modeling of fluid behavior at high pressures becomes the main issue. At this pressure regime, thermodynamic non-idealities should be considered [3, 10]. The use of a real-gas equation of state together with fundamental thermodynamic principles to compute fluid properties is suggested. Although a cubic equation of state may not provide the same accuracy as other methods, its use is recommended because of its easy manageability and its capability to predict the liquid-phase solution. Then, transport properties should be computed using high-pressure models, always ensuring that low-pressure solutions can be recovered. It becomes important not to mix non-ideal and complex models with simpler ones, since better accuracy in these fluid states can be lost. For instance, He et al. [11] show the effects of non-idealities on the diffusion process, where the so-called diffusion barrier can be identified.

Phase-equilibrium laws have to be satisfied at the liquid-gas interface and can be combined with an equation of state to predict mixture composition on each side of the interface. However, works by Dahms and Oefelein suggest that phase-equilibrium results should be carefully used at supercritical pressures [1, 12, 13]. They propose that traditional phase equilibrium can only be applied far enough of the critical point. In the transcritical region, where temperature is approaching its critical value, the liquid-gas interface enters a continuum region. Under this consideration, fluid properties do not experience a discontinuity across the interface, but they evolve continuously. Then, temperature does not become the same on both sides of the interface. Nevertheless, interface thicknesses of the order of a few nanometers are shown, which could be considered as a discontinuity, similar to the practical treatment of shock waves.

The present paper proposes a numerical study of liquid-gas dynamics at high pressures using a non-ideal equation of state and different models to compute transport properties. Particularly, phase-equilibrium laws are imposed at the liquid-gas interface and the development of a diffusion layer on each side is analyzed to infer if a well-established phase equilibrium exists before typical hydrodynamic instabilities, such as the Kelvin-Helmholtz instability, grow sufficiently to disrupt the liquid jet. That is, if fluid properties vary considerably around the interface, the liquid disruption process may be modified.

The rest of this document is structured as follows. First, the problem configuration and the governing equations, as well as different fluid properties models, are presented in Section 2. Then, the numerical methodology to solve the governing equations and the matching conditions at the liquid-gas interface is introduced in Section 3. An analysis of the obtained results, based on reference mixtures of oxygen ($O_2$) and n-octane ($C_8H_{18}$) or oxygen and n-decane ($C_{10}H_{22}$), is shown in Section 4. Finally, a summary of the main findings concludes the paper in Section 5.

## 2. Problem Statement and Governing Equations

*2.1. Problem definition*

A 1-D domain is defined where a liquid composed of pure species B is suddenly introduced into a gas composed of pure species A (see Figure 1). Pressure is constant throughout the domain and temperature is higher in the gas phase, but without exceeding the critical temperature of the pure liquid species. This temperature difference across the interface can drive the vaporization of the liquid phase. Because of phase equilibrium, gas species A will dissolve in the liquid phase and vapor of species B will mix with the gas. This situation will generate diffusion layers composed of both species; that is, a binary mixture will exist on both sides of the interface. To capture these layers, the domain has to be defined large enough to include them at all times. That is, it should become numerically infinite.

When supercritical pressures for the liquid are considered, the mixing of species around the interface allows the coexistence of both phases, as the mixture critical pressure becomes higher than the pure liquid critical pressure and may rise above the chamber pressure.

*2.2. Governing equations*

The governing equations to be solved are the continuity equation, Eq. (1), the species continuity equation, Eq. (2), and the energy equation, Eq. (3). The momentum equation is not needed under the assumptions taken in this and other works [14]. The problem is mainly driven by diffusion forces and fluid velocity becomes very small. Then, momentum flux is negligible compared to the studied high pressures and, together with the constant-pressure assumption, it becomes unnecessary to solve the momentum equation. Only in low-pressure situations this assumption may need to be reconsidered.

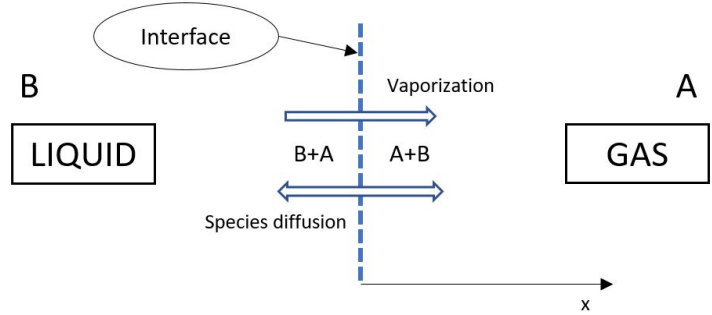

Figure 1: Sketch of the 1-D interface problem.

The one-dimensional equations in Cartesian coordinates are, in both phases,

$$\frac{\partial \rho}{\partial t} + \frac{\partial}{\partial x}(\rho u) = 0 \quad (1)$$

$$\frac{\partial}{\partial t}(\rho Y_i) + \frac{\partial}{\partial x}(\rho u Y_i) + \frac{\partial}{\partial x}(J_i) = 0 \quad (2)$$

$$\frac{\partial}{\partial t}(\rho h) + \frac{\partial}{\partial x}(\rho u h) = \frac{\partial}{\partial x}\left(\lambda \frac{\partial T}{\partial x}\right) - \sum_{i=1}^{N} \frac{\partial}{\partial x}(J_i h_i) \quad (3)$$

where $\rho$, $u$, $Y_i$, $J_i$, $h$, $\lambda$, $T$ and $h_i$ are the mixture density, the fluid velocity, the mass fraction of species $i$, the mass flux of species $i$ due to diffusion, the mixture specific enthalpy, the thermal conductivity, the temperature and the partial specific enthalpy of species $i$. $N$ represents the total number of species. In Eq. (3), the total derivative of pressure and viscous dissipation are neglected, in accordance with the assumptions of this work.

Combining Eq. (1), Eq. (2) and Eq. (3), the non-conservative forms of the species continuity equation, Eq. (4), and the energy equation, Eq. (5), are obtained.

$$\rho \frac{\partial Y_i}{\partial t} + \rho u \frac{\partial Y_i}{\partial x} + \frac{\partial}{\partial x}(J_i) = 0 \quad (4)$$

$$\rho \frac{\partial h}{\partial t} + \rho u \frac{\partial h}{\partial x} - \rho \sum_{i=1}^{N} h_i \frac{\partial Y_i}{\partial t} - \rho u \sum_{i=1}^{N} h_i \frac{\partial Y_i}{\partial x} - \frac{\partial}{\partial x}\left(\lambda \frac{\partial T}{\partial x}\right) + \sum_{i=1}^{N} J_i \frac{\partial h_i}{\partial x} = 0 \quad (5)$$

The solution of these equations, together with matching conditions at the interface, will provide the evolution of mixture properties (e.g., temperature, composition, etc.) around the interface. The numerical treatment of this problem is explained in Section 3.

*2.3. Thermodynamics: Equation of state*

The Soave-Redlich-Kwong [15] cubic equation of state, Eq. (6), has been chosen to predict mixture properties due to its good performance in a wide range of fluid states, including sub-, near- and supercritical regions [16, 17]. Thus, it can represent both gas and liquid phase.

$$p = \frac{R_u T}{v - b} - \frac{a}{v(v + b)} \quad (6)$$

Eq. (6) can be rewritten in the following form

$$Z^3 - Z^2 + (A - B - B^2)Z - AB = 0 \tag{7}$$

where $Z$ is the compressibility factor and $A$ and $B$ are two parameters given by

$$Z = \frac{pv}{R_u T} \quad ; \quad A = \frac{ap}{(R_u T)^2} \quad ; \quad B = \frac{bp}{R_u T} \tag{8}$$

In the previous equations, $p$, $v$ and $R_u$ are the pressure, mixture molar volume and universal gas constant. The temperature-dependent mixture attractive parameter, $a$, and the mixture repulsive parameter, $b$, can be computed using the following mixing rules

$$a = \sum_{i=1}^{N} \sum_{j=1}^{N} X_i X_j (a_i a_j)^{0.5} (1 - k_{ij}) \quad ; \quad b = \sum_{i=1}^{N} X_i b_i \tag{9}$$

where $X_i$ and $k_{ij}$ are the mole fraction of species $i$ and the so-called binary interaction coefficients between species $i$ and $j$. The individual attractive and repulsive parameters can be obtained from the critical point of each species as

$$a_i = 0.42748 \frac{(R_u T_{ci})^2}{p_{ci}} \alpha_i \quad ; \quad b_i = 0.08664 \frac{R_u T_{ci}}{p_{ci}} \tag{10}$$

$$\alpha_i = [1 + S_i (1 - T_{ri}^{0.5})]^2 \quad ; \quad S_i = 0.48508 + 1.55171 \omega_i - 0.15613 \omega_i^2 \tag{11}$$

being $T_{ci}$, $p_{ci}$, $T_{ri} = T/T_{ci}$ and $\omega_i$ the critical temperature, critical pressure, reduced temperature and acentric factor of species $i$.

Solving Eq. (7), the molar volume of the mixture, $v$, can be obtained in both phases and, therefore, mixture density is computed as $\rho = MW/v$, where $MW$ is the molecular weight of the mixture. However, the accuracy of the solutions obtained from this equation of state depend on the binary interaction coefficients, $k_{ij}$. Different methodologies to estimate them are proposed and have been included for the studied mixtures when available [18–22].

Mixture molar enthalpy, $\bar{h}$, and entropy, $\bar{s}$, at high-pressure conditions are computed using fundamental thermodynamic relations. Non-ideal effects are measured by a departure function [23] from ideal-gas values, $\bar{h}^*$ and $\bar{s}^*$, defined in terms of $Z$ as

$$\bar{h} - \bar{h}^* = R_u T(Z - 1) + R_u T \int_{\infty}^{v} \left[ T \left( \frac{\partial Z}{\partial T} \right) \bigg|_v \right] \frac{dv}{v} \tag{12}$$

$$\bar{s} - \bar{s}^* = R_u \ln(Z) + R_u \int_{\infty}^{v} \left[ T \left( \frac{\partial Z}{\partial T} \right) \bigg|_v - 1 + Z \right] \frac{dv}{v} \tag{13}$$

Rearranging Eq. (12) and Eq. (13), using the definitions of $p$ and $Z$ from Eq. (6) and Eq. (8), the specific mixture enthalpy, $h$, and entropy, $s$, are expressed as

$$h = h^*(T) + \frac{1}{MW} \left[ R_u T(Z - 1) + \frac{T(\partial a / \partial T)|_{p,X_i} - a}{b} \ln \left( \frac{Z + B}{Z} \right) \right] \tag{14}$$

$$s = s^*(T, p_0) + \frac{1}{MW} \left[ -R_u \ln \left( \frac{p}{p_0} \right) - R_u \sum_{i}^{N} X_i \ln(X_i) + \frac{1}{b} \left( \frac{\partial a}{\partial T} \right) \bigg|_{p,X_i} \ln \left( \frac{Z + B}{Z} \right) + R_u \ln(Z - B) \right] \tag{15}$$

where $-R_u \ln(p/p_0)$ accounts for the deviation from the pressure reference state of the ideal-gas mixture entropy [24] and $-R_u \sum_i^N X_i \ln(X_i)$ is the so-called entropy of mixing, related to the increase in entropy associated to the mixing process of an ideal-gas mixture [25, 26].

Then, the specific mixture heat capacity at constant pressure, $c_p$, can be obtained by differentiating the specific mixture enthalpy with respect to temperature at constant pressure and composition as

$$c_p = \left(\frac{\partial h}{\partial T}\right)\bigg|_{p,X_i} = c_p^*(T) + \frac{1}{MW}\left[\frac{T}{b}\ln\left(\frac{v+b}{v}\right)\left(\frac{\partial^2 a}{\partial T^2}\right)\bigg|_{p,X_i} - R_u\right] \\ + \frac{1}{MW}\left(\frac{\partial v}{\partial T}\right)\bigg|_{p,X_i}\left[p - \frac{T(\partial a/\partial T)|_{p,X_i} - a}{v(v+b)}\right] \qquad (16)$$

To compute ideal-gas temperature-dependent specific enthalpy, $h^*(T)$, specific entropy, $s^*(T, p_0)$, and specific heat capacity at constant pressure, $c_p^*(T)$, the correlations by Passut and Danner [27] are used, together with ideal mixture rules. For each species $i$, $h_i^*$, $s_i^*$ and $c_{p,i}^*$ become functions of temperature as

$$h_i^*(T) = \hat{A} + \hat{B}T + \hat{C}T^2 + \hat{D}T^3 + \hat{E}T^4 + \hat{F}T^5 \qquad (17)$$

$$s_i^*(T, p_0) = \hat{B}\ln T + 2\hat{C}T + \frac{3}{2}\hat{D}T^2 + \frac{4}{3}\hat{E}T^3 + \frac{5}{4}\hat{F}T^4 + \hat{G} \qquad (18)$$

$$c_{p,i}^*(T) = \hat{B} + 2\hat{C}T + 3\hat{D}T^2 + 4\hat{E}T^3 + 5\hat{F}T^4 \qquad (19)$$

where constants $\hat{A}$-$\hat{G}$ are given for each species and the reference pressure, $p_0$, is set at 1 atm [27]. Then, ideal specific mixture properties are obtained as

$$h^*(T) = \sum_{i=1}^N Y_i h_i^*(T) \quad ; \quad s^*(T, p_0) = \sum_{i=1}^N Y_i s_i^*(T, p_0) \quad ; \quad c_p^*(T) = \sum_{i=1}^N Y_i c_{p,i}^*(T) \qquad (20)$$

Under non-ideal behavior, partial molar enthalpy, $\bar{h}_i$, is obtained from mixture molar enthalpy by applying the definition of partial enthalpy [28]

$$\bar{h}_i = \left(\frac{\partial \bar{h}}{\partial X_i}\right)\bigg|_{p,T,X_{j\neq i}} \qquad (21)$$

which results in

$$\bar{h}_i = \bar{h}_i^*(T) + p\left(\frac{\partial v}{\partial X_i}\right)\bigg|_{p,T,X_{j\neq i}} - R_u T + \frac{aA_1}{v+b}\left[A_2 - \frac{(\partial v/\partial X_i)|_{p,T,X_{j\neq i}}}{v}\right] \\ + \frac{1}{b}\ln\left(\frac{v+b}{v}\right)\left[T\left(\left(\frac{\partial^2 a}{\partial X_i \partial T}\right)\bigg|_{p,X_{j\neq i}} - \left(\frac{\partial a}{\partial X_i}\right)\bigg|_{p,T,X_{j\neq i}}\right) - aA_1 A_2\right] \qquad (22)$$

The parameters $A_1$ and $A_2$ from Eq. (22) are defined as

$$A_1 \equiv \frac{T}{a}\left(\frac{\partial a}{\partial T}\right)\bigg|_{p,X_i} - 1 \quad ; \quad A_2 \equiv \frac{(\partial b/\partial X_i)|_{p,T,X_{j\neq i}}}{b} \qquad (23)$$

Notice that the $R_u T$ term is, in fact,

$$R_u T = R_u T \sum_{i=1}^N X_i \qquad (24)$$

so it still appears in the derivation of Eq. 22.

To obtain the partial specific enthalpy, $h_i$, partial molar enthalpy is divided by the molecular weight of species $i$.

Moreover, fugacity is also computed using the equation of state. It becomes useful to work with the fugacity coefficient, defined for each species $i$ as

$$\Phi_i = \frac{f_i}{p_i} = \frac{f_i}{pX_i} \tag{25}$$

where $f_i$ and $p_i$ are the fugacity and partial pressure of species $i$.

The general thermodynamic relation for the fugacity of a component in a mixture [15] is

$$\ln(\Phi_i) = \int_\infty^v \left[\frac{1}{v} - \frac{1}{R_u T}\left(\frac{dp}{dn_i}\right)\bigg|_{T,p,n_j}\right] dv - \ln Z \tag{26}$$

which again can be rewritten in terms of the equation of state as

$$\ln(\Phi_i) = \frac{b_i}{b}(Z-1) - \ln(Z-B) - \frac{A}{B}\left[2\left(\frac{a_i}{a}\right)^{0.5} - \frac{b_i}{b}\right]\ln\left(1+\frac{B}{Z}\right) \tag{27}$$

*2.4. Interface conservation relations and phase equilibrium*

There are matching conditions that must be satisfied at the liquid-gas interface which establish a connection between the liquid and the gas domains. These relations are based on a mass and energy balance across the interface and thermodynamic equilibrium requirements.

Mass balances across the phase interface are given by Eq. (28) and Eq. (29) and energy balance is given by Eq. (30), considering the relative velocity between the fluid and the interface. The interface velocity is an eigenvalue of the problem and is defined as $U$. The subscripts $l$ and $g$ refer to the liquid phase and the gas phase, respectively.

$$\rho_g(u_g - U) = \rho_l(u_l - U) \tag{28}$$

$$\rho_g Y_{gi}(u_g - U) + J_{gi} = \rho_l Y_{li}(u_l - U) + J_{li} \tag{29}$$

$$\rho_l u_l h_l - \lambda_l \left(\frac{\partial T}{\partial x}\right)_l + \sum_{i=1}^N J_{li} h_{li} - \rho_l U h_l = \rho_g u_g h_g - \lambda_g \left(\frac{\partial T}{\partial x}\right)_g + \sum_{i=1}^N J_{gi} h_{gi} - \rho_g U h_g \tag{30}$$

Eq. (30), in combination with Eq. (28), provide a more manageable energy balance equation,

$$\rho_l(u_l - U)(h_l - h_g) - \lambda_l \left(\frac{\partial T}{\partial x}\right)_l + \sum_{i=1}^N J_{li} h_{li} = -\lambda_g \left(\frac{\partial T}{\partial x}\right)_g + \sum_{i=1}^N J_{gi} h_{gi} \tag{31}$$

Finally, phase-equilibrium relations are given by equality of temperature, pressure and chemical potential of each species on both sides of the interface [9, 23]. The last condition can be rewritten in terms of an equality in fugacity of each species in both phases.

$$p_l = p_g \tag{32}$$

$$T_l = T_g \tag{33}$$

$$f_{li} = f_{gi} \tag{34}$$

If Eq. (34) is combined with Eq. (25), it can be rewritten as

$$X_{li}\Phi_{li} = X_{gi}\Phi_{gi} \tag{35}$$

Eq. (32) is automatically satisfied under the assumptions of uniform pressure and negligible interface curvature. The jump in pressure that could exist across a curved liquid-gas interface due to surface tension is given by

$$\frac{\Delta p}{p} = \frac{\sigma}{\kappa p} \tag{36}$$

where $\kappa$ and $\sigma$ are the radius of curvature and the surface tension of the interface, respectively.

Using the Macleod-Sugden correlation [23], estimates of the surface tension using interface properties from the results of Section 4.4 provide $\sigma = 5.031 \times 10^{-3}$ N/m at 10 bar and $\sigma = 1.931 \times 10^{-3}$ N/m at 150 bar. For $\kappa = 1$ $\mu$m, which represents the typical smallest values that could be found in liquid breakup problems, the jumps in pressure across the interface represent 0.5% and 0.013% of the chamber pressure for the 10 bar and the 150 bar cases, respectively. Therefore, even when the interface curvature should not be neglected, the jump in pressure becomes negligible for the phase equilibrium analysis as the chamber pressure is increased, while it will still be important for the breakup analysis.

*2.5. Transport properties*

Thermal conductivity, $\lambda$, is computed using the correlations from Chung et al. [29], which can also be used to compute viscosity, $\mu$, and can be extended to mixtures. Then, diffusion mass fluxes, $J_i$, are computed using the generalized Maxwell-Stefan equations, Eq. (37), which allow the inclusion of non-ideal effects in the diffusion process [11, 14, 30].

$$\sum_{j \neq i} \frac{X_i X_j}{D_{ij}} \left( \frac{J_j^M}{cX_j} - \frac{J_i^M}{cX_i} \right) = d_i - \sum_{j \neq i} \frac{X_i X_j}{D_{ij}} \left( \frac{D_j^T}{\rho Y_j} - \frac{D_i^T}{\rho Y_i} \right) \nabla \ln T \tag{37}$$

The driving force for species $i$, $d_i$, is defined as

$$d_i = \nabla X_i + X_i \sum_{j=1}^{N} \left( \frac{\partial \ln \Phi_i}{\partial X_j} \right)_{p,T} \nabla X_j \tag{38}$$

where non-idealities are introduced in the fugacity coefficient term, which can be obtained from differentiation of Eq. (27).

Because temperature gradients are small in our problem, Eq. (37) can be rewritten neglecting thermal diffusion effects. For a binary mixture in a Cartesian 1-D configuration, it becomes

$$\frac{1}{c} \frac{X_1 X_2}{D_{12}} \left( \frac{J_2^M}{X_2} - \frac{J_1^M}{X_1} \right) = \Gamma_{12} \frac{\partial X_1}{\partial x} \tag{39}$$

where $J_i^M$, $c$, $D_{12}$ and $\Gamma_{12}$ are the mole-based diffusion flux, the molar density, the binary diffusion coefficient and the thermodynamic factor expressing non-ideal effects, defined as

$$\Gamma_{12} = 1 + X_1 \left[ \left( \frac{\partial \ln \Phi_1}{\partial X_1} \right)_{p,T} - \left( \frac{\partial \ln \Phi_1}{\partial X_2} \right)_{p,T} \right] \tag{40}$$

The binary diffusion coefficient, $D_{12}$, is computed using Leahy-Dios and Firoozabadi correlations [30]. Eq. (39), together with the property $\sum_i^N J_i^M = 0$, provides the solution for the diffusion fluxes. However, the governing equations are solved using mass-based diffusion fluxes, $J_i^m$ or $J_i$, which can be obtained from mole-based diffusion fluxes by applying a transformation or change of reference frame [30, 31].

## 3. Numerical Method

### 3.1. Discretization of the equations

Eqs. (1), (4) and (5) are discretized using finite-volume techniques [32]. An explicit first-order time-integration scheme is used instead of high-order explicit approaches or implicit methods [14] to discretize Eqs. (4) and (5). Since we are interested in the early stages of flow evolution, the small time step restriction does not become a critical point of the numerical solution. Furthermore, the explicit treatment simplifies the coupling between the different equations. However, the selected time step must satisfy the Courant-Friedrichs-Lewy (CFL) conditions to ensure numerical stability [14, 33]. To discretize Eq. (1), an implicit scheme is used to obtain the updated velocity field from density changes throughout the domain.

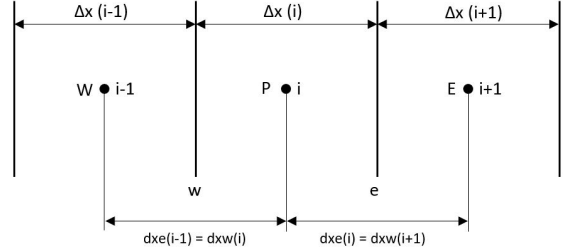

Figure 2: Mesh definition of the 1-D domain.

Following the mesh arrangement shown in Figure 2, the discretized governing equations are

$$\frac{\rho_P^{n+1} - \rho_P^n}{\Delta t} V_P + [(\rho u)_e^{n+1} - (\rho u)_w^{n+1}] \Delta y \Delta z = 0 \tag{41}$$

$$\rho_P^n \frac{Y_{P,i}^{n+1} - Y_{P,i}^n}{\Delta t} V_P + (\rho_P u_P)^n [Y_{e,i}^n - Y_{w,i}^n] \Delta y \Delta z + [J_{e,i}^n - J_{w,i}^n] \Delta y \Delta z = 0 \tag{42}$$

$$\rho_P^n \frac{h_P^{n+1} - h_P^n}{\Delta t} V_P + (\rho_P u_P)^n [h_e^n - h_w^n] \Delta y \Delta z - \rho_P^n \sum_{i=1}^{N} h_{P,i}^n \frac{Y_{P,i}^{n+1} - Y_{P,i}^n}{\Delta t} V_P$$

$$- (\rho_P u_P)^n \sum_{i=1}^{N} h_{P,i}^n [Y_{e,i}^n - Y_{w,i}^n] \Delta y \Delta z - \left[ \left( \lambda \frac{\partial T}{\partial x} \right)_e^n - \left( \lambda \frac{\partial T}{\partial x} \right)_w^n \right] \Delta y \Delta z \tag{43}$$

$$+ \sum_{i=1}^{N} J_{P,i}^n [h_{e,i}^n - h_{w,i}^n] \Delta y \Delta z = 0$$

where the subscript $P$ refers to the node we are solving for and the subscripts $e$ and $w$ refer to the cell faces east and west from node $P$. The superscripts $n + 1$ and $n$ refer to the new and old time step evaluation, respectively. Moreover, the grid parameters $V_P$, $\Delta y$ and $\Delta z$ are the cell volume, the cell size in the vertical direction and the cell size in the spanwise direction, respectively. For the 1-D treatment, $\Delta y = 1$ and $\Delta z = 1$. Therefore, $V_P = \Delta x \Delta y \Delta z = \Delta x$.

Velocity and diffusion mass fluxes are evaluated at the cell faces, while all other properties, such as mass fractions or density, are evaluated at the nodes. A second-order central-difference scheme is used to compute these variables at the node or at the cell face. Since small velocities are found in this problem, using central differences does not introduce any numerical instabilities and produces similar results as upwind schemes [14]. For example,

$$u_P = \frac{1}{2}(u_e + u_w) \quad ; \quad Y_{w,i} = \frac{1}{2}(Y_{P,i} + Y_{W,i}) \tag{44}$$

Gradients at the cell faces are computed following this example for temperature:

$$\left(\frac{\partial T}{\partial x}\right)_e = \frac{T_E - T_P}{\text{dxe}} \quad ; \quad \left(\frac{\partial T}{\partial x}\right)_w = \frac{T_P - T_W}{\text{dxw}} \tag{45}$$

However, one issue arises when computing the gradients at the interface. Since a discontinuity is present, central differences would produce erroneous values. Thus, the gradient can only be calculated from node values on one side, either for the liquid phase or the gas phase. To compute it more accurately, a one-sided second-order Taylor series expansion is used, defined accordingly to the parameters shown in Figure 3. Eqs. (46) and (47) show the computation of the temperature gradients on the two sides using this method.

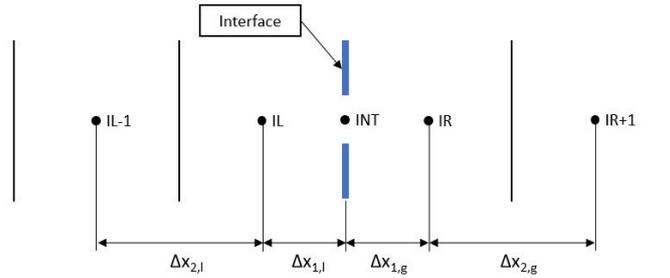

Figure 3: Mesh definitions for the computation of gradients at the interface.

$$\left(\frac{\partial T}{\partial x}\right)_g = \frac{T(IR+1) - T_{int} + \frac{T_{int}-T(IR)}{(\Delta x_{1,g})^2}(\Delta x_{1,g} + \Delta x_{2,g})^2}{\left[(\Delta x_{1,g} + \Delta x_{2,g}) - \frac{1}{\Delta x_{1,g}}(\Delta x_{1,g} + \Delta x_{2,g})^2\right]} \tag{46}$$

$$\left(\frac{\partial T}{\partial x}\right)_l = \frac{T(IL-1) - T_{int} + \frac{T_{int}-T(IL)}{(\Delta x_{1,l})^2}(-\Delta x_{1,l} - \Delta x_{2,l})^2}{\left[-(\Delta x_{1,l} + \Delta x_{2,l}) + \frac{1}{\Delta x_{1,l}}(-\Delta x_{1,l} - \Delta x_{2,l})^2\right]} \tag{47}$$

*3.2. Soltuion algorithm*

The problem is approached by working with a velocity field relative to a fixed interface. That is, working with $U = 0$. In this scenario, fluid will be flowing in or out of the domain, with an associated momentum flux. However, since velocities are small, this momentum flux becomes negligible compared to the high-pressures being analyzed. Therefore, assuming a fixed interface without balancing the momentum fluxes on both domain ends is a reasonable approach. The simulations prove that momentum flux becomes rapidly negligible compared to pressure after a few time steps.

The matching conditions from Section 2.4, Eqs. (28)-(35), the evaluation of temperature gradients at the interface, Eqs. (46) and (47), $\sum_{i=1}^{N} X_{li} = 0$ and $\sum_{i=1}^{N} X_{gi} = 0$ form a closed system of $2N + 8$ equations when working with the relative fluid velocities at the interface or a fixed interface. The number of equations to be solved can further be reduced to $2N + 6$ equations since Eq. (32) is automatically satisfied and the summation over $N$ of Eq. (29) results in Eq. (28).

For a binary mixture, phase equilibrium only provides two points in the thermodynamic space, thus becoming decoupled from the diffusion mass fluxes. For mixtures with $N > 2$, other solution algorithms exist to find the correct phase equilibrium results under specific restrictions [9, 14]. To solve this system of equations, the energy balance equation is defined as the objective function to solve for in terms of the interface temperature, $f(T_{int})$. Given a guessed interface temperature, phase-equilibrium equations provide mixture compositions on both sides of the interface, which are used to evaluate other fluid properties. Then, mass flux balances provide values for the fluid velocity. Knowing these estimates, the objective function is solved using a root-finding algorithm.

The solution of the governing equations is obtained in two different subdomains: the liquid phase and the gas phase. The interface and the ends of the numerical domain act as boundaries for each subdomain, imposing the corresponding boundary conditions. For Eqs. (42) and (43), matching conditions at the interface impose mass fractions and temperature, while far-end boundaries impose constant temperature and pure species concentration for each phase. Velocities computed at the interface act as boundary conditions for Eq. (41).

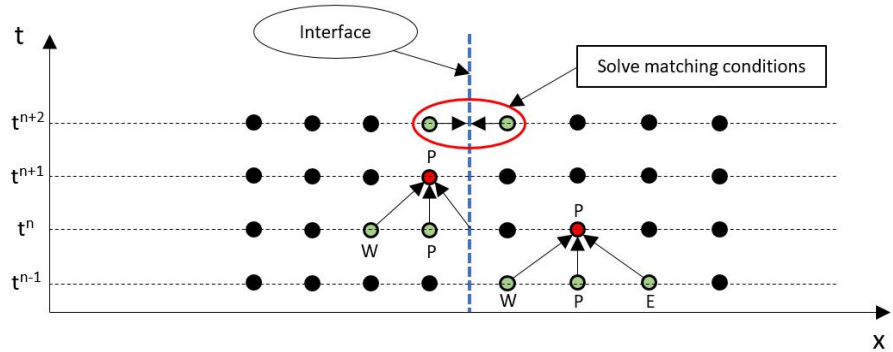

Figure 4: Scheme of the temporal integration procedure.

The temporal integration procedure is shown in Figure 4. To solve the equations, the explicit scheme uses the values computed in the previous time step for the node of interest, $P$, and its neighbors, $E$ and $W$. If $P$ lies immediately on either side of the interface, the interface will act as a neighboring point.

The algorithm is set as follows. First, Eq. (42) is solved, obtaining mass fractions. Then, Eq. (43) is addressed to find new values for the mixture enthalpy. A root-finding algorithm is used to obtain the fluid temperature according to its composition and enthalpy by using Eq. (14). To evaluate mixture density, the equation of state is used. Once all fluid properties are known, the matching conditions at the interface are solved and new velocities are obtained, which are used to solve Eq. (41) and obtain the velocity field.

Grid independence studies over a total simulation time of 100 $\mu$s with a mixture of $O_2$ and $C_{10}H_{22}$ show that, for the 10-bar case, reference interface values and fluid properties are accurately computed with a uniform grid spacing, $\Delta x$, of 100 nm and a time step, $\Delta t$, of 0.5 ns. For 50-150 bar, a grid spacing of 200 nm and a time step of 5 ns are sufficient. A domain length, $L$, of 400 $\mu$m with the interface located at 100 $\mu$m from the liquid end of the domain is large enough to include both diffusion layers during the elapsed time at all pressures.

## 4. Results and Discussion

### 4.1. Phase equilibrium

The first issue to be addressed is the corroboration that two-phase behavior can be sustained at supercritical pressures with the methodology implemented in this work. Mixtures of light gases with heavy hydrocarbons have been tested to prove that this coexistence occurs. Figure 5 shows mixture compositions obtained with the Soave-Redlich-Kwong equation of state together with phase-equilibrium equations for combinations of oxygen, $O_2$, and nitrogen, $N_2$, with n-octane, $C_8H_{18}$, and n-decane, $C_{10}H_{22}$. Table 1 shows the molecular weight, critical properties and acentric factor of these species. In the plots of Figure 5, the mole fraction of the fuel (i.e., the heavy hydrocarbon) is given for the liquid and the gas phase as a function of chamber pressure (in terms of the reduced pressure of the fuel, $p_r = p/p_{c,fuel}$) and temperature.

| Species | MW (kg/mol) | $p_c$ (bar) | $T_c$ (K) | $\omega$ |
|---|---|---|---|---|
| $O_2$ | 0.031999 | 50.43 | 154.58 | 0.0222 |
| $N_2$ | 0.028013 | 33.96 | 126.19 | 0.0372 |
| $C_8H_{18}$ | 0.114230 | 24.97 | 569.32 | 0.3950 |
| $C_{10}H_{22}$ | 0.142280 | 21.03 | 617.70 | 0.4884 |

Table 1: Properties of the species used in this work.

The results prove that two-phase behavior can be predicted at pressures higher than the critical pressure of the injected fuel, similar to other works implementing the Redlich-Kwong, the Soave-Redlich-Kwong or the Peng-Robinson equations of state [9, 11, 14, 34–37]. As pressure is increased, the dissolution of the oxidizer (i.e., light gas) in the liquid phase is enhanced. The two-phase behavior can be maintained as this dissolution generates a liquid solution close to the interface with different critical properties than the pure liquid species. Section 4.2 discusses the evaluation of these mixture critical properties.

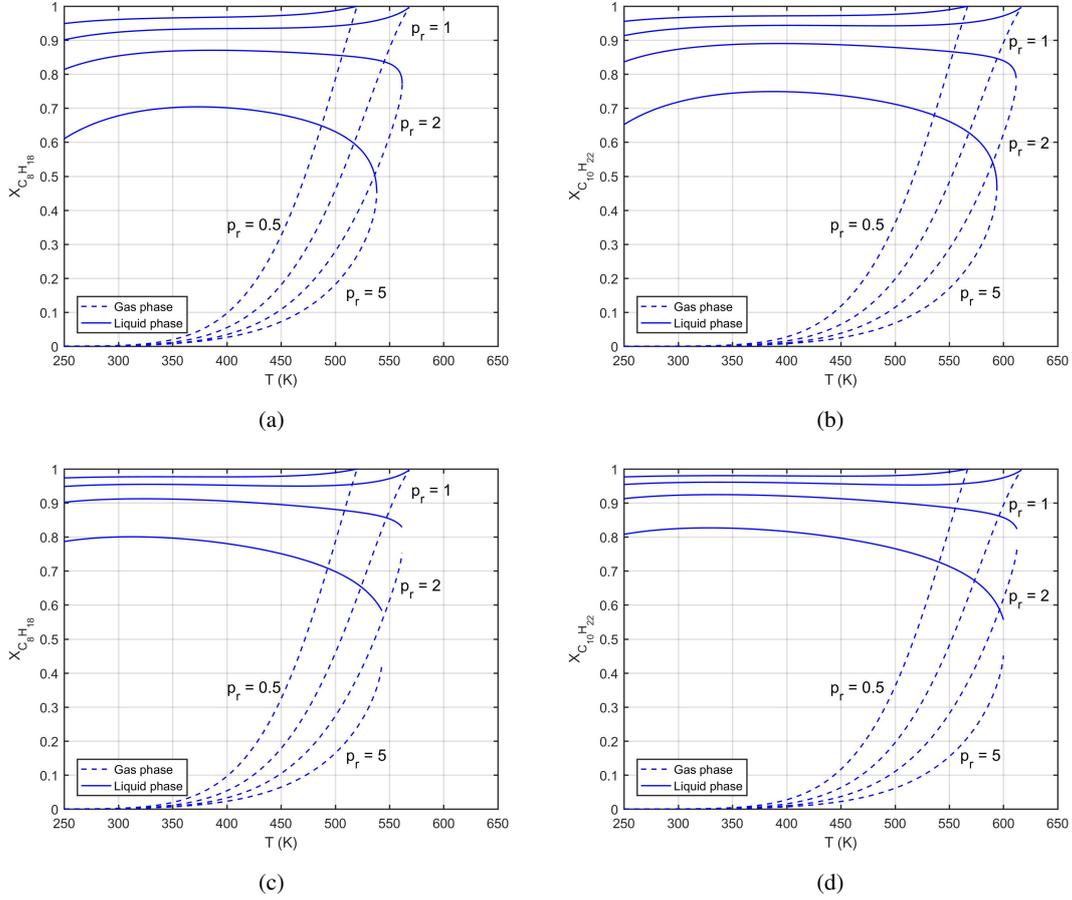

Figure 5: Phase equilibrium results for different pairs of species obtained using the Soave-Redlich-Kwong equation of state to predict fugacity coefficients. (a) $O_2$ / $C_8H_{18}$; (b) $O_2$ / $C_{10}H_{22}$; (c) $N_2$ / $C_8H_{18}$; (d) $N_2$ / $C_{10}H_{22}$.

## *4.2. Diffusion layer development*

A problem using liquid n-decane and gas oxygen has been chosen as a reference case to analyze the physical implications of liquid injection at high pressures. $C_{10}H_{22}$ is introduced at 450 K and $O_2$ at 550 K, thus below the critical temperature of n-decane. Four different pressures have been analyzed: a subcritical 10-bar case and three supercritical scenarios at 50, 100 and 150 bar.

The effects of increasing pressure on mixture composition and density are seen in Figure 6, which provides field results evaluated at $t = 100\ \mu s$. The dissolution of $O_2$ in the liquid phase is enhanced as pressure increases because of phase-equilibrium requirements at the interface, while $C_{10}H_{22}$ mixing in the gas phase is reduced. The development of a diffusion layer around the interface can clearly be seen, where compressibility effects strongly modify mixture density as pressure increases. For the low-pressure case, a diffusion layer affecting fluid density is observed, but providing a nearly incompressible behavior.

The diffusion layer thickness slightly increases in the liquid phase as pressure is increased, mainly be-cause of the enhanced dissolution of the lighter species. On the other hand, the thickness of the diffusion layer is reduced in the gas phase. That is, the higher gas density reduces the ability of n-decane molecules to diffuse through the gas phase. However, high-pressure situations create diffusion layers where density varies more sharply than low-pressure situations.

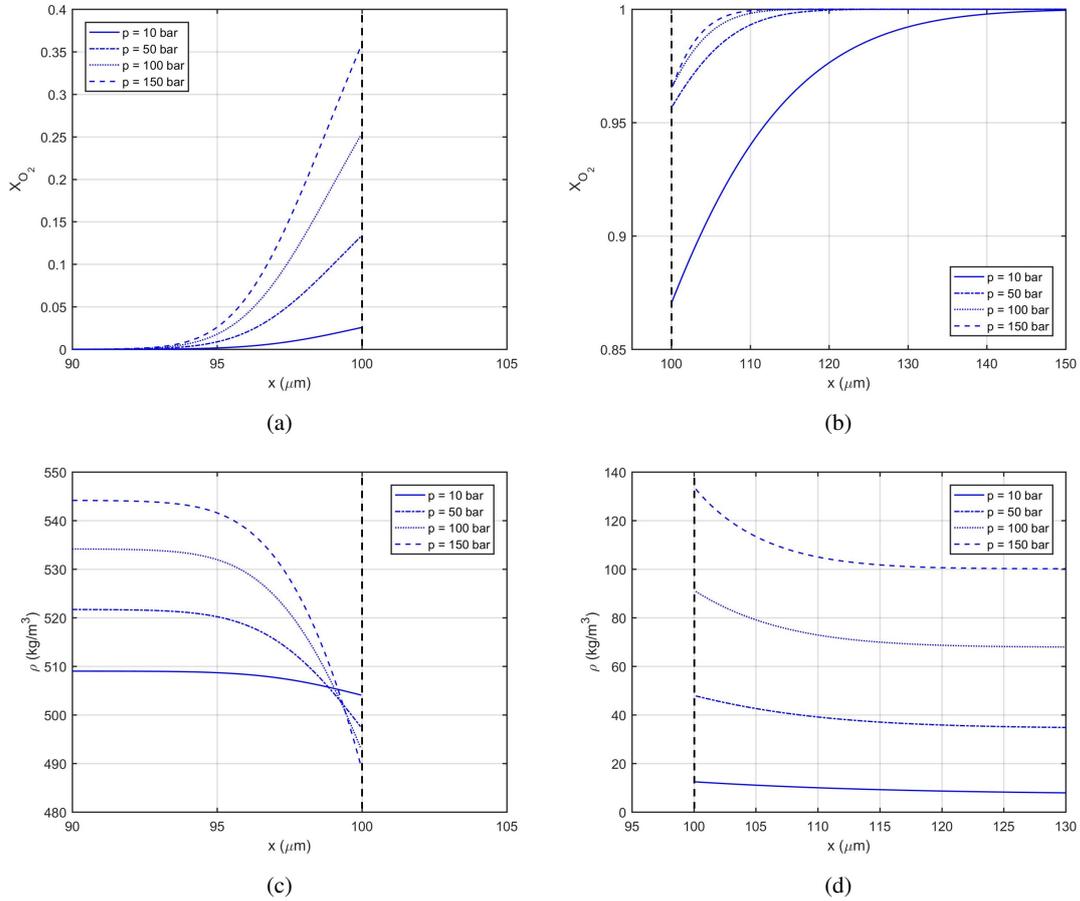

Figure 6: Pressure effects on oxygen mole fraction and fluid density on each side of the interface for the oxygen/n-decane mixture at t = 100 $\mu$s. (a) $O_2$ liquid mole fraction; (b) $O_2$ gas mole fraction; (c) liquid density; (d) gas density.

Figure 7 shows the temporal evolution of temperature and density profiles near the interface for 150 bar in order to determine how fast these diffusion layers grow. Notice that, due to the configuration of the problem, interface properties tend to reach steady values during the elapsed simulation time. As seen in Figure 8, interface temperature reaches steady values for times below 10 $\mu$s. Other fluid properties, such as density, take longer to reach a steady behavior. It is observed that, at times between 10-100 $\mu$s, mass and thermal diffusion layers present thicknesses of the order of 3-10 $\mu$m in the liquid phase and 10-30 $\mu$m in the gas phase. For lower pressures, the gas mixture presents thicker layers in the same temporal range. Other tested mixtures (e.g., oxygen/n-octane, nitrogen/n-octane and nitrogen/n-decane) present similar qualitative

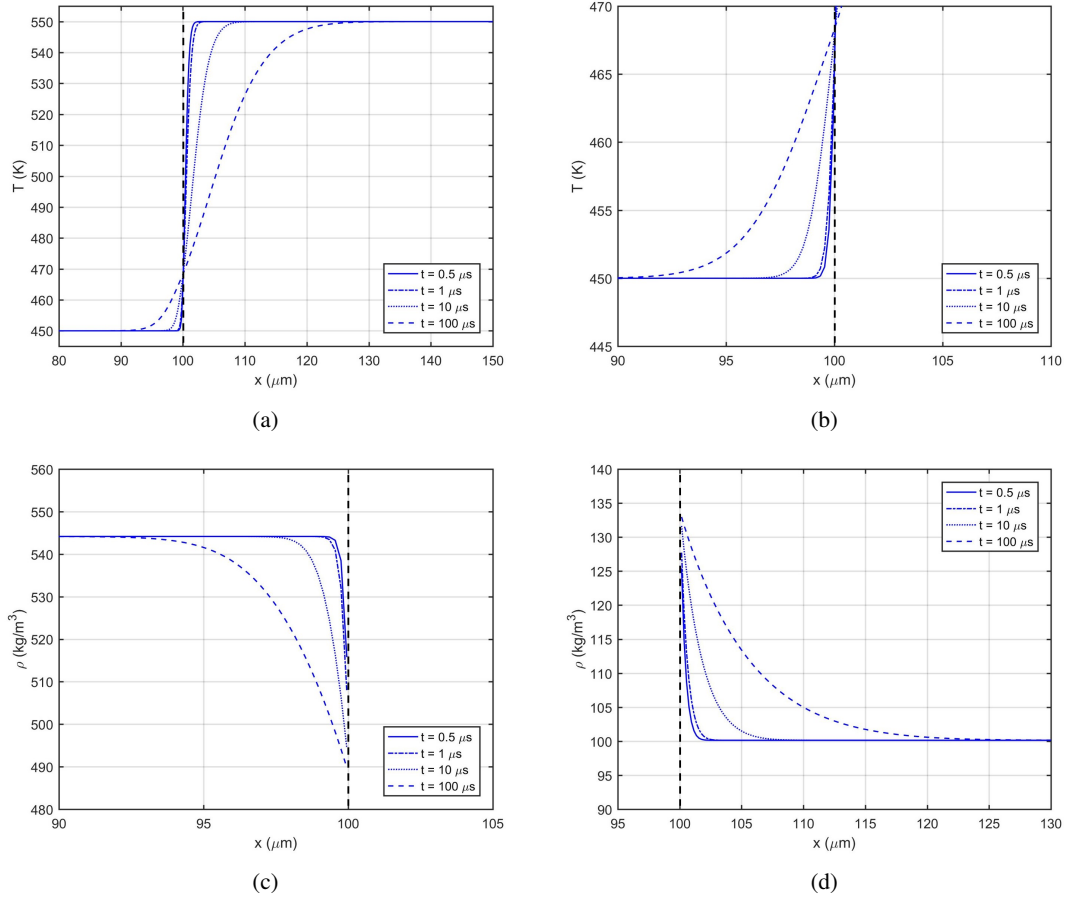

Figure 7: Temporal evolution of the temperature distribution and the fluid density for the oxygen/n-decane mixture at p = 150 bar. (a) temperature; (b) temperature in the liquid phase near the interface; (c) liquid density; (d) gas density.

results and diffusion layer thicknesses of the same order of magnitude are observed.

The temporal evolution of the different diffusion layers seen in Figure 7 suggests that some similarity solution may exist. The thermal diffusion layers have been analyzed by tracking the location $x^*$ as the position where $|T - T_{ambient}| = \frac{1}{10}|T_{int} - T_{ambient}|$, where $T_{ambient}$ is the temperature of the fluid at the far end of the liquid or gas domain. Then, the temporal evolution of the similarity variable $\xi = |x^* - x_{int}|/\sqrt{\alpha_m t}$ is plotted in Figure 9, where $\alpha_m = \frac{\lambda}{\rho c_p}$ is the thermal diffusion coefficient of the mixture at the interface and $x_{int}$ the location of the interface. For long simulation times, once the interface properties are approaching steadiness, it is observed that the thermal diffusion layer evolution could be expressed in terms of only one variable, $\xi$. Further analyses show that enthalpy evolution also presents similar behavior with the temperature evolution, thus a similarity could also be obtained after some transient.

It has been observed that a well-defined two-phase behavior can be established at supercritical pressures for the pure liquid species. However, it still needs to be clarified if the newly-formed liquid solution is subcritical close to the interface for the analyzed pressures. As recommended by Poling et al. [23], an estimation of mixture true critical properties can be obtained from Chueh and Prausnitz [38], where correlations based on experimental data are proposed to obtain critical temperature and critical volume, while a modified Redlich-Kwong equation of state is used to estimate critical pressure. All three computations require the evaluation of some binary interaction coefficients. To compute critical temperature and volume, the methodology proposed by Fenghour [39] can be used to estimate those coefficients if no experimental data is available. The coefficients to evaluate critical pressure can be obtained from vapor-liquid equilibrium experimental data, similar to the $k_{ij}$ defined previously in Section 2.3. A good matching between our phase-equilibrium diagrams and the predicted critical points is observed in many of the tested cases. The present work uses the Soave-Redlich-Kwong equation of state, which is also a modification of the original Redlich-Kwong equation of state.

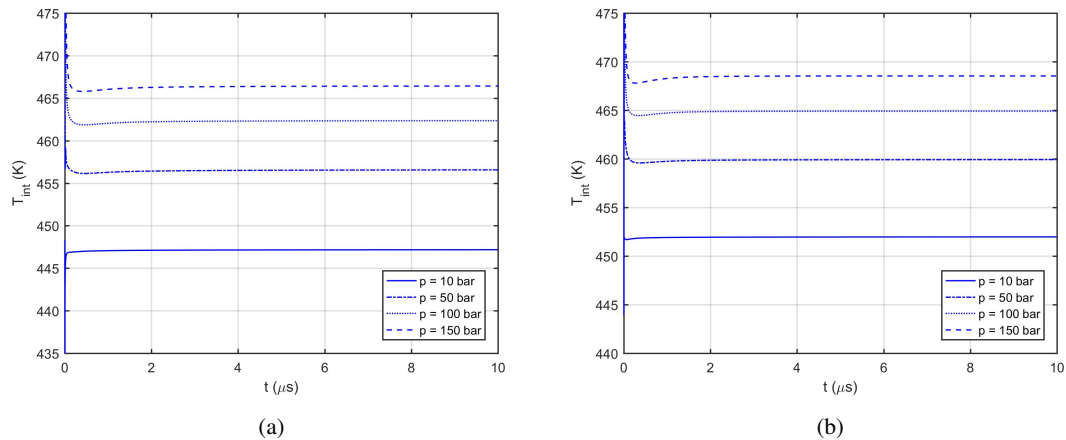

Figure 8: Temporal evolution of the interface temperature for different pressures and mixtures. (a) $O_2$ / $C_8H_{18}$; (b) $O_2$ / $C_{10}H_{22}$.

Figure 10 shows results obtained by applying the methodology from [38, 39] for both oxygen/n-octane and oxygen/n-decane mixtures. Critical pressure shows a peak often reported in other works [38, 40], but usually not rising more than one order of magnitude above the critical pressures of the pure fluids. Thus, it is suspected that the results provided in the range of $X_{O_2} = 0.75 - 0.95$ might not be accurate for the analyzed mixtures, which show peaks two orders of magnitude larger. This issue is corroborated in Figure 11.

Figure 11 shows how the mixture critical points are obtained, using as an example the oxygen/n-octane mixture. Interface equilibrium results obtained from the simulations provide liquid and gas phase compositions at the analyzed pressures. The vertical red line is the interface temperature, while the horizontal red lines represent the mole fractions of n-octane at the interface, both in the liquid and in the gas phase. To locate the critical point of the liquid solution at the interface, the composition is fixed at the liquid equilibrium value. Then, it is sufficient to find the new equilibrium curve with a maximum in temperature at that composition (i.e., the point where both liquid phase and gas phase have the same composition). This point in the phase-equilibrium diagram provides information about both the critical pressure and critical temperature of the liquid solution. The same procedure can be applied to obtain the critical point of the gas mixture. In general, given any mixture composition, its critical properties can be obtained following the same analysis, which becomes a graphical interpretation of the analytical methodology to obtain the critical points of a mixture [41].

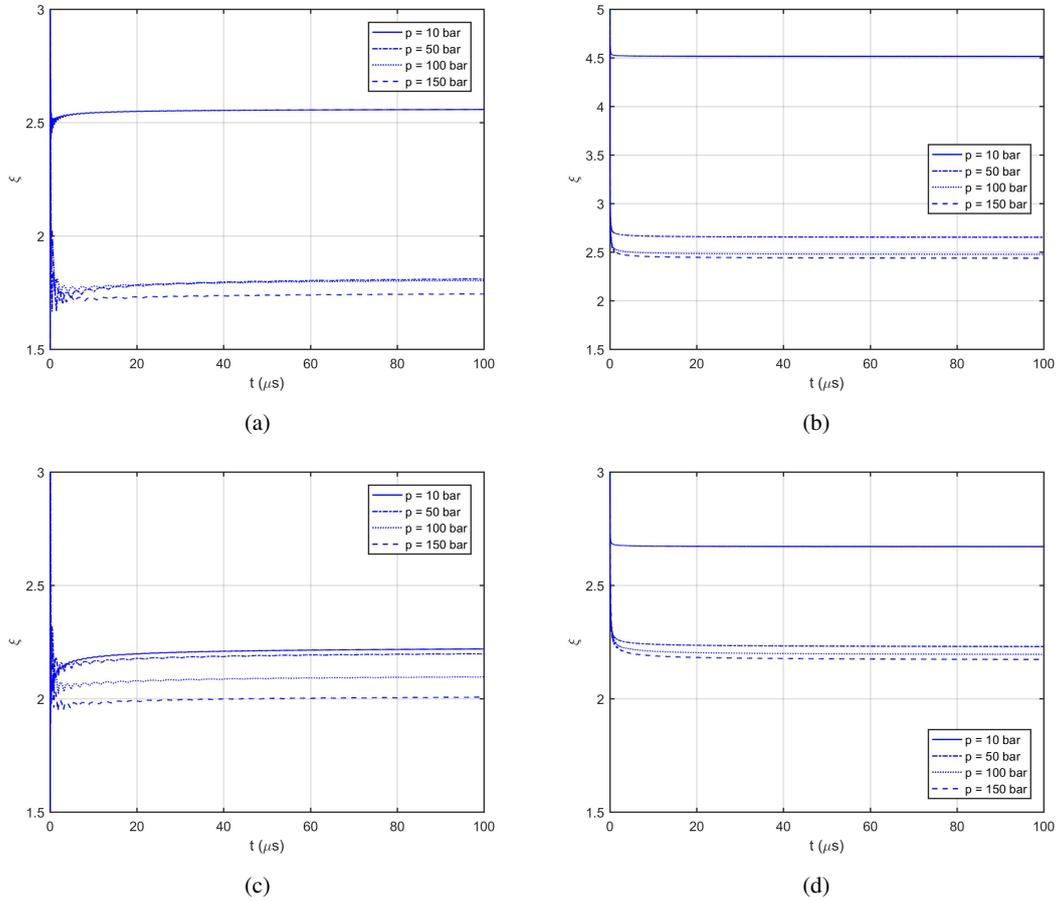

Figure 9: Similarity analysis of the temporal evolution of the temperature profile for different pressures and mixtures. (a) liquid phase $O_2$ / $C_8H_{18}$; (b) gas phase $O_2$ / $C_8H_{18}$; (c) liquid phase $O_2$ / $C_{10}H_{22}$; (d) gas phase $O_2$ / $C_{10}H_{22}$.

For the 10-bar case (Figure 11a), the predicted critical pressures using Chueh and Prausnitz methodology [38] reasonably provide pressure curves in the phase-equilibrium diagram which have a temperature maximum in the desired composition. However, for the 50-bar case and above (Figure 11b), only estimations of the liquid phase critical pressure from Chueh and Prausnitz method seem to be in accordance with the methodology developed in this article. For the gas phase, the predicted critical pressure is 817.5 bar, for which no equilibrium diagram can be found. Nevertheless, if the pressure curve matching the gas phase composition is drawn, a more reasonable critical pressure of 485 bar is obtained. In all analyzed cases, deviations between the obtained critical temperature and the predicted critical temperature, as well as deviations in critical pressure, increase as pressure is increased. These difficulties in predicting the critical properties of a fluid mixture arise from the lack of available experimental data to predict accurate binary interaction coefficients, $k_{ij}$, and also cause question about the validity of simple mixing rules for certain types of mixtures or in certain composition ranges.

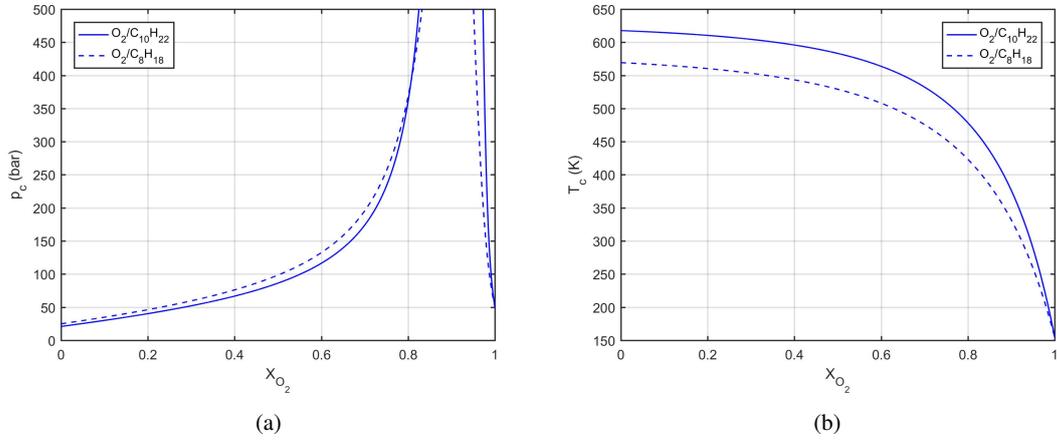

Figure 10: Estimation of mixture critical properties for oxygen/n-octane and oxygen/n-decane as a function of composition. (a) critical pressure; (b) critical temperature.

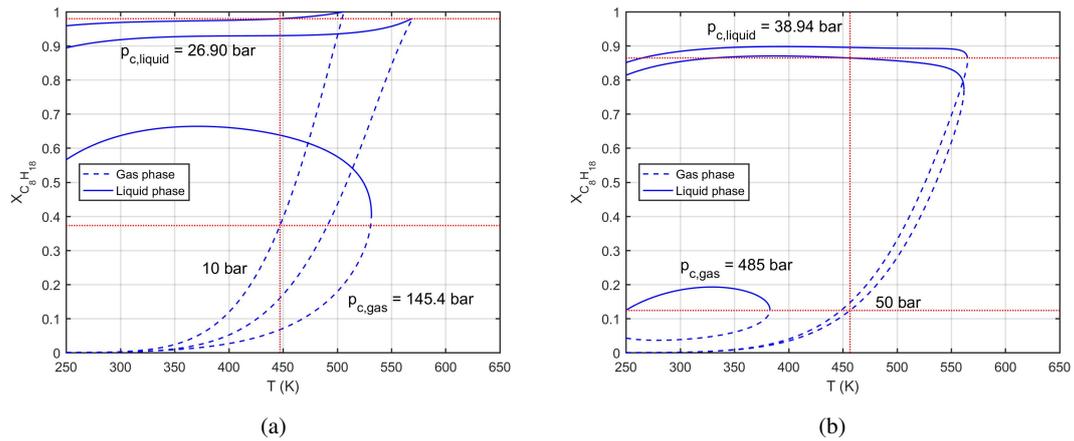

Figure 11: Estimation of mixture critical properties for oxygen/n-octane phase-equilibrium results at different pressures. (a) 10 bar; (b) 50 bar, with self-computed gas phase critical pressure.

An important insight can be extracted from these results. As it can be inferred from Figure 11b, our methodology predicts that at supercritical pressures for the pure liquid species (50-, 100- and 150-bar cases), the liquid solution at the interface is still supercritical. Thus, it can be better described as a compressible liquid. However, two-phase behavior can be sustained. The dissolution of oxygen into the liquid phase increases the critical pressure of the liquid solution, but it does not exceed the chamber pressure. The shape of the phase-equilibrium diagram not only depends on the equation of state being used to model it, but also on the species composing the fluid mixture. Under some configurations, it is possible that the mole fraction of the pure liquid species reaches a maximum in the liquid-phase equilibrium composition as temperature increases for a fixed pressure, just to drop again down to the critical point. Other authors report similar behaviors, also from experimental data points [9, 11–15, 22, 34–37].

When matching conditions at the interface result in a liquid solution composition with higher hydrocarbon mole fraction than the critical mixture composition at that pressure, the resulting critical pressure of the liquid solution lies below the chamber pressure. That is, the analyzed pressure is still supercritical for the liquid phase. Nevertheless, equilibrium conditions (i.e., equality in chemical potentials of both phases) are satisfied and a clear two-phase behavior with a defined liquid-gas interface can exist.

*4.3. Thermodynamic analysis*

The results show that very high pressures have an important effect on the liquid-gas interface dynamics. As seen in Figure 12, where the plots of the fluid velocity are shown for both oxygen/n-octane and oxygen/n-decane mixtures as a function of chamber pressure, the problem transitions from a vaporization to a condensation phenomenon at the interface. That is, the velocity behavior is reversed (i.e., the velocity changes sign). Similar results can be observed in Figure 13, where interface temperature, liquid density, gas density and velocity (computed as if the liquid at the end of the domain were sitting on a wall with zero velocity) are plotted.

In this scenario, condensation by increasing pressure is achieved, even at a supercritical pressure environment. This phenomenon reversal occurs due to mixing of species by diffusion. While heat still conducts from the hotter gas to the colder liquid, the energy flux summed from heat conduction and energy transport by mass diffusion reverses. For low-pressure situations, the internal energy decreases across the interface from the gas to the liquid phase, thus causing vaporization. However, for sufficiently high pressure environments, the internal energy increases across the interface and condensation occurs. Gas-mixture enthalpy is always higher than liquid-solution enthalpy at the interface, while internal energy of the liquid phase exceeds internal energy of the gas phase for high-pressure situations. That is, as pressure is increased and the enthalpy of vaporization is reduced, pressure effects have a stronger impact on the $p/\rho$ term from $\hat{u} = h - p/\rho$, where here $\hat{u}$ is the mixture specific internal energy. Then, gas-mixture internal energy can drop below the liquid-solution internal energy at the interface. Figures 12 and 13 suggest that this phenomenon is more easily achieved for mixtures with less volatile hydrocarbons. For oxygen/n-octane, condensation is only achieved for pressures higher than 100 bar. For oxygen/n-decane, condensation would occur for pressures slightly higher than 50 bar.

The interface velocity plot from Figure 13 shows that, as pressure is increased, the liquid phase goes from contracting to expanding, even before condensation is observed. For low pressures (i.e., below 30 bar for oxygen/n-decane), the low dissolution rate of oxygen into the liquid phase and the vaporization occurring at the interface translate to a reduction of the liquid-solution volume. For higher pressures (i.e., between 30-50 bar for oxygen/n-decane), the dissolution of lighter species into the liquid generates a region of higher specific volume close to the interface, which can overcome the vaporization rate and increase the volume occupied by the liquid. Then, for the same analyzed mixture, pressures above 50 bar show an increase of the liquid-phase volume due to the dissolution of oxygen and the condensation at the interface. Similar conclusions can be extracted for the oxygen/n-octane mixture.

Condensation occurs globally as the net value of mass flux (i.e., summation of hydrocarbon and oxygen mass fluxes) changes sign as pressure increases. Nevertheless, since n-decane or n-octane is being diffused into the gas phase, the hydrocarbon should always vaporize. Similarly, oxygen is diffusing into the liquid phase and should be condensing at all pressures. As shown in Figure 14, the combined convection and diffusion mass fluxes show that the hydrocarbon always vaporizes and oxygen always condenses. That is, the hydrocarbon presents a positive mass flux which relates to the species flowing from the liquid phase into the interface and oxygen presents negative mass flux, which means it is flowing from the gas phase into the interface. Only at higher pressures, as the dissolution of oxygen is increased, the combination of hydrocarbon and oxygen mass fluxes will become negative.

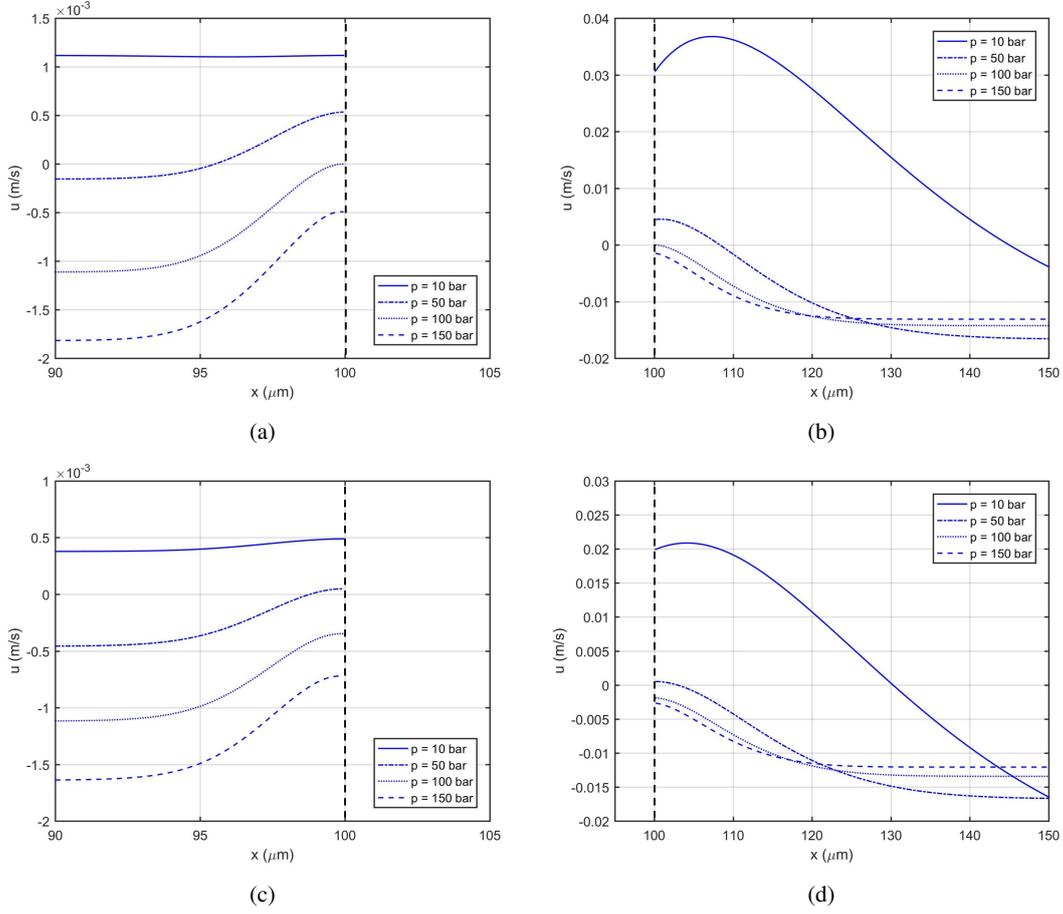

Figure 12: Pressure effects on the velocity field for different mixtures at t = 100 $\mu$s. (a) liquid phase $O_2$ / $C_8H_{18}$; (b) gas phase $O_2$ / $C_8H_{18}$; (c) liquid phase $O_2$ / $C_{10}H_{22}$; (d) gas phase $O_2$ / $C_{10}H_{22}$.

To check the viability of the physical processes observed in the simulations, the First and Second Law of thermodynamics have been applied. Taking a fixed-mass element containing both thermal and mass diffusion layers, its internal energy, $\hat{U}$, enthalpy, $H$, and entropy, $S$, are defined as

$$\hat{U} \equiv \int \hat{u} dm = \int \rho \hat{u} dx \quad ; \quad H \equiv \int h dm = \int \rho h dx \quad ; \quad S \equiv \int s dm = \int \rho s dx \tag{48}$$

where $dm = \rho dx$ is the differential of mass and $\hat{u}$, $h$ and $s$ are the specific internal energy, enthalpy and entropy, as defined previously in this article.

Under this configuration, the First Law is reduced to $\Delta \hat{U} = W$, since negligible heat flux or mass flux is crossing the boundaries of the mass element, thus $Q = 0$. Moreover, kinetic energy is also neglected under the low-speed flow conditions of the problem. That is, the increase over time in internal energy of the mass element, $\Delta \hat{U}$, is directly related to the work done on the mass element, $W$. Furthermore, under the constant-pressure process, the temporal variation in enthalpy becomes $\Delta H = Q|_p = 0$. Therefore, $\Delta \hat{U} = -p\Delta V = W$. To compute $\hat{U}$, $H$ and $S$ at each instant of time, the instantaneous specific internal energy, enthalpy and entropy are integrated following Eq. (48). Figures 15-17 address these considerations for the oxygen/n-decane mixture.

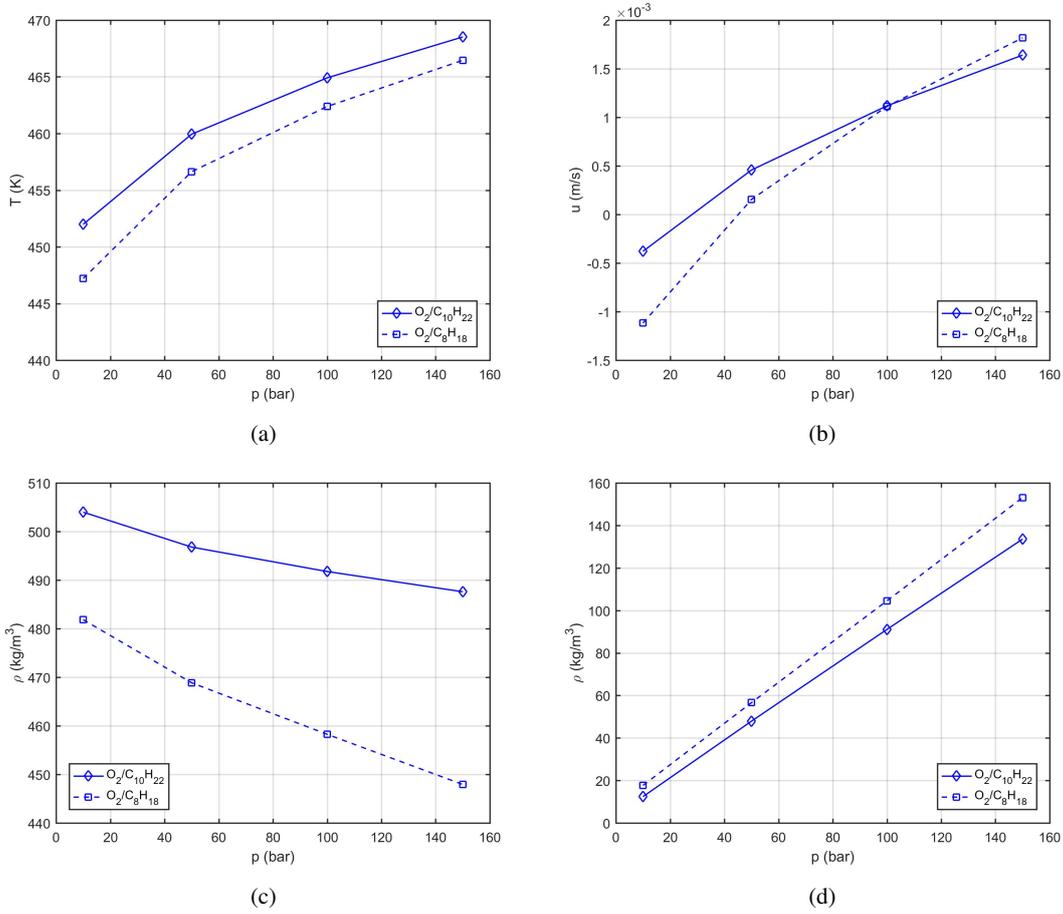

Figure 13: Pressure effects on the interface properties for oxygen/n-octane and oxygen/n-decane mixtures at t = 100 $\mu$s. (a) temperature; (b) velocity; (c) liquid density; (d) gas density.

In Figure 15, the First Law is seen to be satisfied: the integration over time of the work done on the mass element is equal to the increase in internal energy of that mass. A small deviation appears related to numerical errors. For the mesh being used, the error at 100 $\mu$s represents about 0.63-1.21 % of the increase in internal energy. Further refinement of the mesh reduces these errors.

The combined First and Second Law for a constant-pressure process must also be satisfied (i.e., $\Delta H = \int dH = \iint \frac{dh}{dt} dm dt = \iint T \frac{ds}{dt} dm dt$). Due to the difficulty of evaluating the integral over time of the right hand side term, the validity of this relation is checked by computing the local error from time step to time

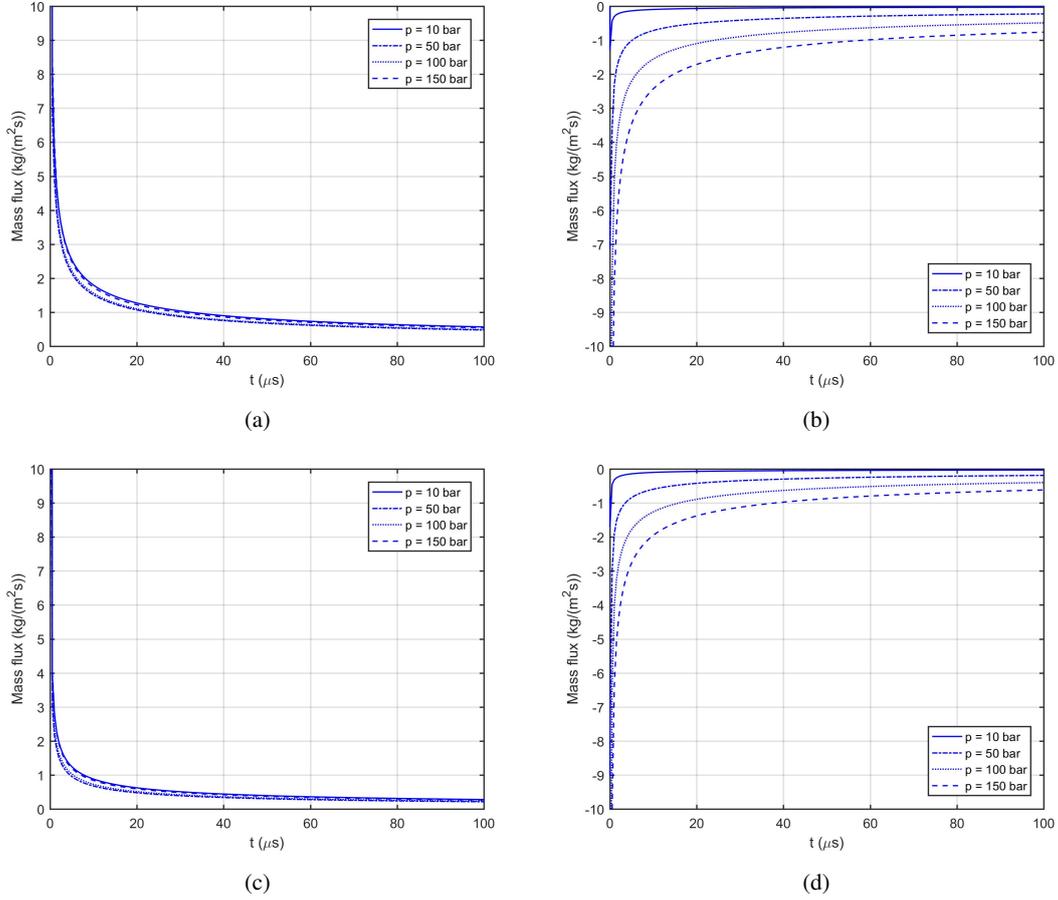

Figure 14: Pressure effects on individual mass fluxes for different mixtures. (a) $O_2$ / $C_8H_{18}$ octane mass flux; (b) $O_2$ / $C_8H_{18}$ oxygen mass flux; (c) $O_2$ / $C_{10}H_{22}$ decane mass flux; (d) $O_2$ / $C_{10}H_{22}$ oxygen mass flux

step of $dh = Tds$ and integrating these errors over the mass element. The computed errors are of the order of $10^{-2}$-$10^{-3}$ at the initial time steps, since larger gradients are present. Then, those errors rapidly decrease as time progresses. Refinement in the temporal scale of the mesh further reduces the magnitude of these errors.

The implications of $\Delta H = 0$ are seen in Figure 16. For the 10-bar case, enthalpy does not change considerably in the liquid phase. The enthalpy distribution in the gas phase presents a wavy shape, which allows for $\Delta H$ to become zero. Analysis of the energy equation terms shows that conduction due to temperature difference across the interface causes an energy sink in the gas phase. After an initial period, the diffusion and vaporization of n-decane into the gas, with the higher associated enthalpy to that species, acts as an energy source able to raise again the enthalpy of the gas mixture. On the other hand, the 150-bar case shows that enthalpy is conserved as it increases in the liquid phase and drops in the gas phase. In this scenario, the energy introduced into the gas mixture by the diffusion of n-decane cannot overcome the sink caused by heat conduction and reversed advection (i.e., condensation).

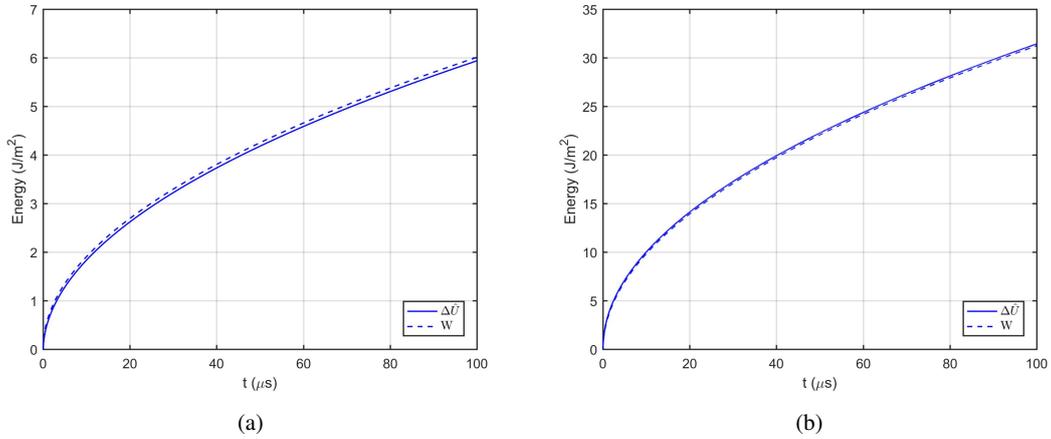

Figure 15: Validation of the First Law for the oxygen/n-decane mixture. (a) 10 bar; (b) 150 bar.

The specific entropy distribution is given in Figure 17. Entropy increases in the liquid phase and decreases in the gas mixture, in accordance with the temperature behavior. The entropy of the mass element increases with time (i.e., $\Delta S = \int dS = \iint \frac{ds}{dt} dm dt > 0$), as seen in Figure 18. Since no heat diffuses into the system and viscous effects are negligible, the increase in entropy is directly related to the irreversibility of the mixing process within the mass element.

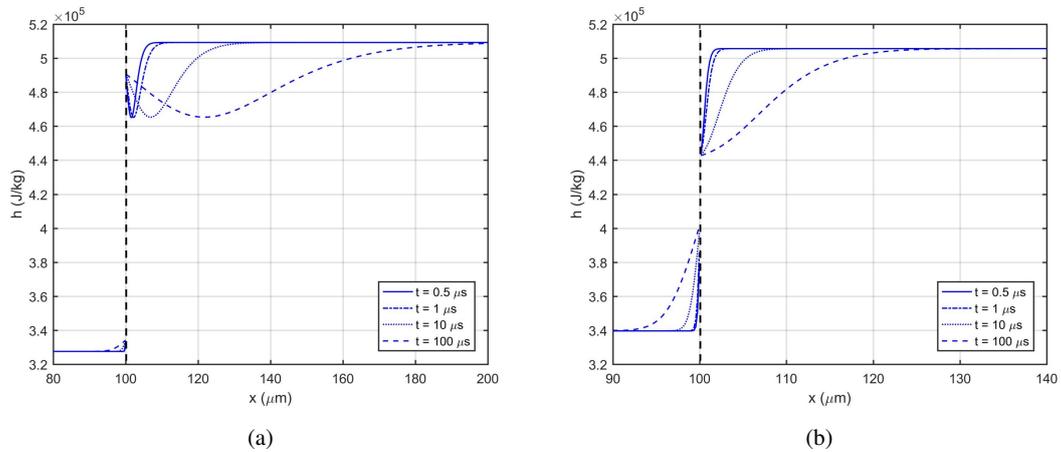

Figure 16: Temporal evolution of specific enthalpy of the oxygen/n-decane mixture. (a) 10 bar; (b) 150 bar.

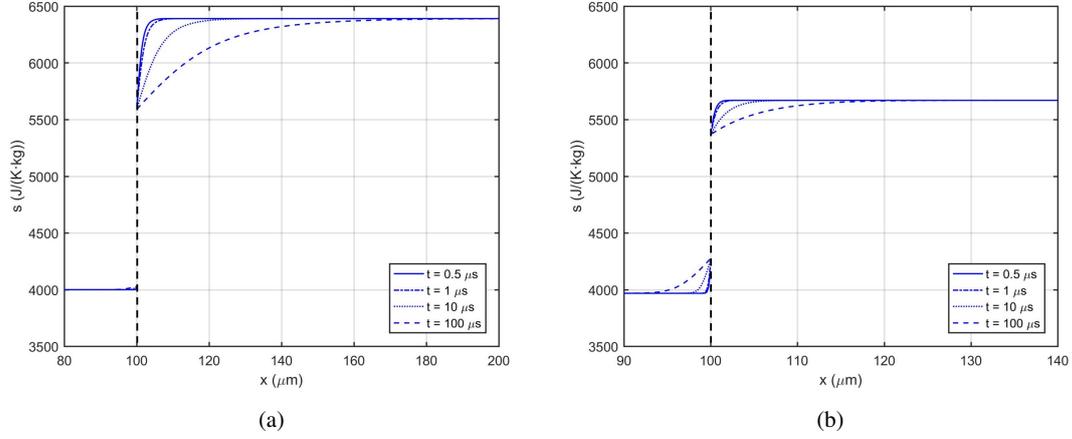

Figure 17: Temporal evolution of specific entropy of the oxygen/n-decane mixture. (a) 10 bar; (b) 150 bar.

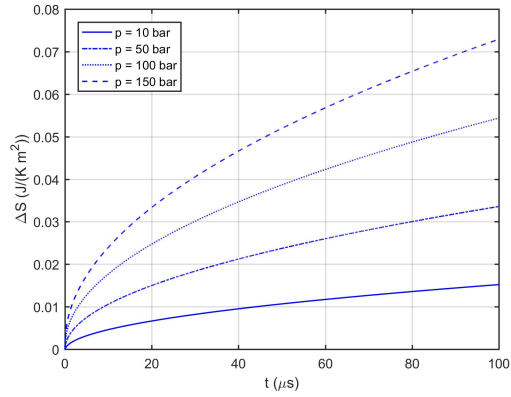

Figure 18: Temporal evolution of the variation of entropy in the mass element.

*4.4. Implications for the hydrodynamic instabilities*

Kelvin-Helmholtz (KH) hydrodynamic instabilities are analyzed to determine the possible effects that high-pressure environments may have on the liquid disruption process. We make a rough comparison between the reciprocal of the KH growth rates for two parallel incompressible flows with interface surface tension and the time scales developed earlier in this article. The issue is whether clear two-phase behavior is established before substantial growth in KH waves occur. As a first approximation, perturbations at the liquid-gas interface for liquid sheets flowing parallel to a gas can be analyzed by studying the evolution of the perturbation amplitude [42, 43]. It is given by

$$\eta(x,t) = \hat{\eta} e^{ct} e^{ikx} \tag{49}$$

where $\eta$ represents the perturbation amplitude. It will depend on the oscillation amplitude, $\hat{\eta}$, growth rate, $c$, time, $t$, wave number, $k = 2\pi/\lambda$, and location, $x$.

Particularly interesting is the effect of the growth rate parameter, $c$, whose real part determines if a perturbation is stable ($Re(c) < 0$) or unstable ($Re(c) > 0$). Linearizing the problem, an expression for $c$ as a function of the fluid properties is found to be [43]

$$c = -i\frac{k(\rho_g u_g + \rho_l u_l)}{\rho_g + \rho_l} - k^2\frac{\mu_g + \mu_l}{\rho_g + \rho_l} \pm \left[\frac{\rho_g \rho_l k^2 (u_g - u_l)^2}{(\rho_g + \rho_l)^2} - \frac{(\rho_l - \rho_g)gk}{\rho_g + \rho_l} \right. \\ \left. - \frac{\sigma k^3}{\rho_g + \rho_l} + \frac{k^4(\mu_g + \mu_l)^2}{(\rho_g + \rho_l)^2} + 2ik^3\frac{(\rho_g \mu_l - \rho_l \mu_g)(u_g - u_l)}{(\rho_g + \rho_l)^2}\right]^{1/2} \quad (50)$$

which includes viscosity, $\mu$, and surface tension, $\sigma$, effects. However, this expression only considers viscous potential flow. Thus, no vorticity or shear stress exists and only normal viscous stress appears. Viscosity at the interface is obtained using Chung et al. correlations [29], as explained in Section 2.5. Then, surface tension is again computed using the Macleod-Sugden correlation, as recommended by Poling et al. [23].

Different situations combining chamber pressures and injection velocities have been analyzed using the reference mixture of oxygen/n-decane. Table 2 provides information about the studied cases and Table 3 provides numerical values for the properties of interest, which are the most unstable wavelength, $\lambda_w$, the real part of the growth rate, $c_R = Re(c)$, the characteristic time for the perturbation to grow by a factor of $e$, $\tau$, and the perturbation frequency, $f$.

| Study case | Pressure (bar) | Injection velocity (m/s) |
|---|---|---|
| A | 10 | 10 |
| B | 10 | 100 |
| C | 150 | 10 |
| D | 150 | 100 |

Table 2: Analyzed cases for the Kelvin-Helmholtz instability.

| Study case | $\lambda_w$ ($\mu$m) | $c_R$ (1/s) | $\tau$ ($\mu$s) | $f$ (1/s) |
|---|---|---|---|---|
| A | 40 | $1.368 \times 10^5$ | 7.310 | $2.423 \times 10^5$ |
| B | 0.39 | $1.128 \times 10^8$ | $8.865 \times 10^{-3}$ | $2.375 \times 10^8$ |
| C | 1.95 | $7.175 \times 10^6$ | 0.139 | $3.930 \times 10^6$ |
| D | 0.032 | $3.196 \times 10^9$ | $3.129 \times 10^{-4}$ | $2.260 \times 10^9$ |

Table 3: Results of Kelvin-Helmholtz instability for the oxygen/n-decane mixture.

The numerical results provided in Table 3 show that both the increase in pressure and the increase in injection velocity reduce the wavelengths of the most unstable perturbations and increase their growth rate. This confirms that high-pressure injection scenarios with low injection velocities could present similar behavior as high-speed injection at low pressures.

The increase in relative velocity between the liquid sheet and the gas allows for inertia effects to easily disrupt the liquid stream. Furthermore, when pressure is increased, gas and liquid density are seen to be more alike as the critical point of the mixture is approached. This similarity in densities causes a reduction of the surface tension holding the liquid together, as well as an increased inertia of the gas around the liquid. Therefore, both effects will cause a fast disruption of the liquid jet.

However, the numerical results provided in Table 3 should only be considered qualitatively. Eq. (50) does not include shear-layer effects, which will damp the perturbations (i.e., decrease the growth rates) and move the most unstable waves to higher wavelengths. Moreover, the effect of variable density around the interface is not included. Actually, under high-pressure situations, the unstable waves will present wavelengths between 1-100 $\mu$m and growth rates around 20-100 $\mu$s [5–8], which correspond to substantial temporal evolution of the diffusion layer growth studied in this work.

## 5. Summary and Conclusions

A methodology to compute supercritical fluid behavior has been developed considering a non-ideal equation of state and various models to describe fluid properties and liquid-gas interactions. It has been proven that phase-equilibrium conditions enhance the dissolution of gas species into the liquid phase, obtaining a mixture close to the liquid-gas interface with different critical properties. The liquid phase presents higher critical pressure than the pure liquid species, but still below the chamber pressure for the high-pressure cases. However, equilibrium conditions allow the coexistence of two phases.

Under these conditions, the development of mass diffusion layers on each side of the interface is analyzed to obtain the time scale of this process, as well as its characteristics. It is observed that high pressures allow diffusion layer thicknesses of 3-10 $\mu$m in the liquid solution in the temporal range of 10-100 $\mu$s. On the other hand, diffusion layer thickness in the gas mixture is reduced as gas density increases and layer thicknesses of 10-30 $\mu$m are observed. Density variations within the diffusion layers become more relevant as pressure increases, showing that not only a formation of a diffusion layer appears, but also fluid properties change more abruptly across it.

Furthermore, it has been seen that increasing pressure allows for a typically vaporization-driven problem at low pressures to present condensation at higher pressures, while still satisfying the laws of thermodynamics. Thus, condensation may occur even when heat conducts from the hotter gas to the colder liquid. Taking a fixed-mass element containing all diffusion layers, the increase in internal energy is directly related to the work done on the element. Globally, enthalpy is conserved and entropy increases over time due to the irreversibility of the mixing process.

A comparison with preliminary results of typical hydrodynamic instabilities confirms that high-pressure effects will cause a similar liquid disruption as in high-speed liquid injection problems. Comparing with information available in the literature, it is expected that phase equilibrium around the interface will be well established before hydrodynamic instabilities disrupt the liquid, since similar time scales are present. Nevertheless, more realistic simulations need to be done to include all the real-fluid effects on high-pressure injection problems (e.g., viscous layers and shear stress), by extending the problem configuration to a 2-D or 3-D domain where the injection of a liquid jet into a gas is analyzed.

## Acknowledgments

The authors would like to thank Prof. Ahmed F. Ghoniem and his students Ping He and Ashwin Raghavan of the Massachusetts Institute of Technology for useful discussions. This research has been partially funded by a Balsells Graduate Fellowship.